\documentclass[12pt,preprint]{aastex}

\usepackage{rotating}

\usepackage{booktabs,threeparttable}

\slugcomment{To appear in Apj}
\shorttitle{BLR Micro and millilensing}
\shortauthors{Guerras et al.}

\begin{document}

\title{Microlensing of Quasar Broad Emission Lines: Constraints on Broad Line Region Size}


\author{E. GUERRAS\altaffilmark{1,2}, E. MEDIAVILLA\altaffilmark{1,2}, J. JIMENEZ-VICENTE\altaffilmark{3,4}, C.S. KOCHANEK\altaffilmark{5}, J. A. MU\~NOZ\altaffilmark{6}, E.
FALCO\altaffilmark{7}, V. MOTTA\altaffilmark{8}}

\altaffiltext{1}{Instituto de Astrof\'{\i}sica de Canarias, V\'{\i}a L\'actea S/N, La Laguna 38200, Tenerife, Spain}
\altaffiltext{2}{Departamento de Astrof\'{\i}sica, Universidad de la Laguna, La Laguna 38200, Tenerife, Spain}
\altaffiltext{3}{Departamento de F\'{\i}sica Te\'orica y del Cosmos, Universidad de Granada, Campus de Fuentenueva, 18071 Granada, Spain}
\altaffiltext{4}{Instituto Carlos I de F\'{\i}sica Te\'orica y Computacional, Universidad de Granada, 18071 Granada, Spain}
\altaffiltext{5}{Department of Astronomy and the Center for Cosmology and Astroparticle Physics, The Ohio State University, 4055 McPherson Lab, 140 West 18th Avenue, Columbus, OH, 43221 }
\altaffiltext{6}{Departamento de Astronom\'{\i}a y Astrof\'{\i}sica, Universidad de Valencia, 46100 Burjassot, Valencia, Spain.}
\altaffiltext{7}{Center for Astrophysics, 60 Garden Street, Cambridge, MA 02138, USA}
\altaffiltext{8}{Departamento de F\'{\i}sica y Astronom\'{\i}a, Universidad de Valpara\'{\i}so, Avda. Gran Breta\~na 1111, Valpara\'{\i}so, Chile}

\begin{abstract}
We measure the differential microlensing of the broad emission lines 
between 18 quasar image pairs in 16 gravitational lenses.  We find
{that the broad emission lines are in general weakly microlensed}. {The results show, at a modest level of confidence ($ 1.8\sigma$),} that high ionization lines such as CIV are more strongly microlensed
than low ionization lines such as H$\beta$, indicating that the high
ionization line emission regions are more compact. If we statistically model
the distribution of microlensing magnifications, we obtain estimates
for the broad line region size of $r_s=24_{-15}^{+22}$ and $r_s=55_{-35}^{+150}$~light-days
(90\% confidence) for the high and low ionization lines, respectively.
When the {samples are divided into higher and lower luminosity quasars}, we find that the line emission regions of more luminous quasars are larger, with a slope consistent with the
expected scaling from photoionization models. Our estimates also agree well with the results from local reveberation 
mapping studies.
\end{abstract}



\section{Introduction}

Strong, broad emission lines are characteristic of many active galactic nuclei (AGN), and their physical origins are important by virtue of their proximity to the central
engine and their potential use as probes of the gas flows either fueling the AGN or feeding mass and energy back into the host galaxy.  To date, the primary probe of the
geometry and kinematics of the broad line regions has been reverberation mapping,  where the delayed response of the emission line flux to changes in the photoionizing
continuum is used to estimate the distance of the line emitting material from the central engine  (see, e.g., the  reviews by Peterson 1993, 2006).   Reverberation mapping
studies have shown that the global structure of the broad line region is consistent with photoionization models, with the  radius increasing with the (roughly) square root of
the continuum luminosity (e.g. \citealt{Bentz2009}) and high ionization lines (e.g. CIV) originating at smaller radii than low ionization lines (e.g. H$\beta$).  Recent
studies have increasingly focused on measuring the delays as a function of line velocity in order to understand the kinematics of the broad line region (Denney et al. 2009,
2010, Bentz et al. 2010, Brewer et al. 2011, Doroshenko et al. 2012, Pancoast et al. 2012).  The results to date suggest that there is no common kinematic structure, with
differing sources showing signs of inward, outward and disk-like velocity structures.

While very successful, reverberation mapping suffers from several limitations.  First,
the studies are largely limited to relatively nearby, lower luminosity AGN because the
delay time scales for distant, luminous quasars are longer than existing monitoring
programs can be sustained.  Not only do the higher luminosities increase the intrinsic
length of the delay (which is then further lengthened by the cosmological redshift),
but the higher luminosity quasars also have lower variability amplitudes (see, e.g.,
\citealt{Macleod2010}).  Second, one of the most important applications of the results of reverberation 
mapping at present is as a calibrator for estimating black hole masses from single
epoch spectra (\citealt{Wandel1999}).  These calibrations are virtually all for the H$\alpha$ and H$\beta$
lines, while the easiest lines to measure for high redshift quasars are the MgII
and CIV lines because the Balmer lines now lie in the infrared.  Without direct
calibrations, there is a contentious debate about the reliability of MgII 
(e.g., \cite{Mclure2002}, \cite{Kollmeier2006}, \cite{Shen2008}, \cite{Onken2008})
and CIV (e.g., \cite{Vestergaard2006}, \cite{Netzer2007}, \cite{Fine2010}, \cite{Assef2011}) 
black hole mass estimates.  

An alternative means of studying the structure of the broad line region is to 
examine how it is microlensed in gravitationally lensed quasars.  In microlensing,
the stars in the lens galaxy differentially magnify components of the quasar
emission regions leading to time and wavelength dependent changes in the flux
ratios of the images (see the review by Wambsganss 2006).  The amplitude of the
magnification is controlled by the size of the emission region, with smaller
source regions showing larger magnifications. The broad line region was
initially considered to be too large to be affected by microlensing 
(Nemiroff 1988, Schneider \& Wambsganss 1990), but for sizes consistent with
the reverberation mapping results the broad line regions should
show microlensing variability (see \citealt{Mosquera2011}) as explored in theoretical studies by 
Abajas et al. (2002, 2007), Lewis \& Ibata (2004) and Garsden et al (2011). 
Observational evidence for microlensing in the broad line region has been discussed for Q2237+0305 (Lewis
et al. 1998, Metcalf et al. 2004, Wayth et al. 2005, 
Eigenbrod et al. 2008, O'Dowd et al. 2010, Sluse et al. 2011), SDSS J1004+4112 (Richards et al. 2004, 
G\'omez-\'Alvarez et al. 2006, Lamer et al. 2006, Abajas et al. 2007) and 
SDSS J0924+0219 (Keeton et al. 2006), as well as in broader surveys by Sluse et al. (2012)
and Motta et al. (2012).  For example, in their detailed study of 
Q2237+0305, Sluse et al. (2011) demonstrated the power of microlensing, obtaining estimates 
of the BLR size for both { CIII] ($r_{CIII]}\sim 49_{-35}^{+103}$~light-days) and CIV  ($r_{CIV}\sim 66_{-46}^{+110}$~light-days) }
emission lines.  Like reverberation mapping, the microlensing size estimates can
also be made as a function of velocity, and the two methods can even be combined
to provide even more detailed constraints (see Garsden et al. 2011). 

Here we survey microlensing of the broad emission lines in a sample of {18} pairs of
lensed quasar images compiled by Mediavilla et al. (2009).  In \S2 we describe
the data and show that the line core and higher velocity wings are 
differentially microlensed.  In \S3 we use these differences to 
derive constraints on the size of the line emitting regions and we summarize
the results in \S4.

\section{Data Analysis \label{DA}}

\begin{figure}[h]
\plotone{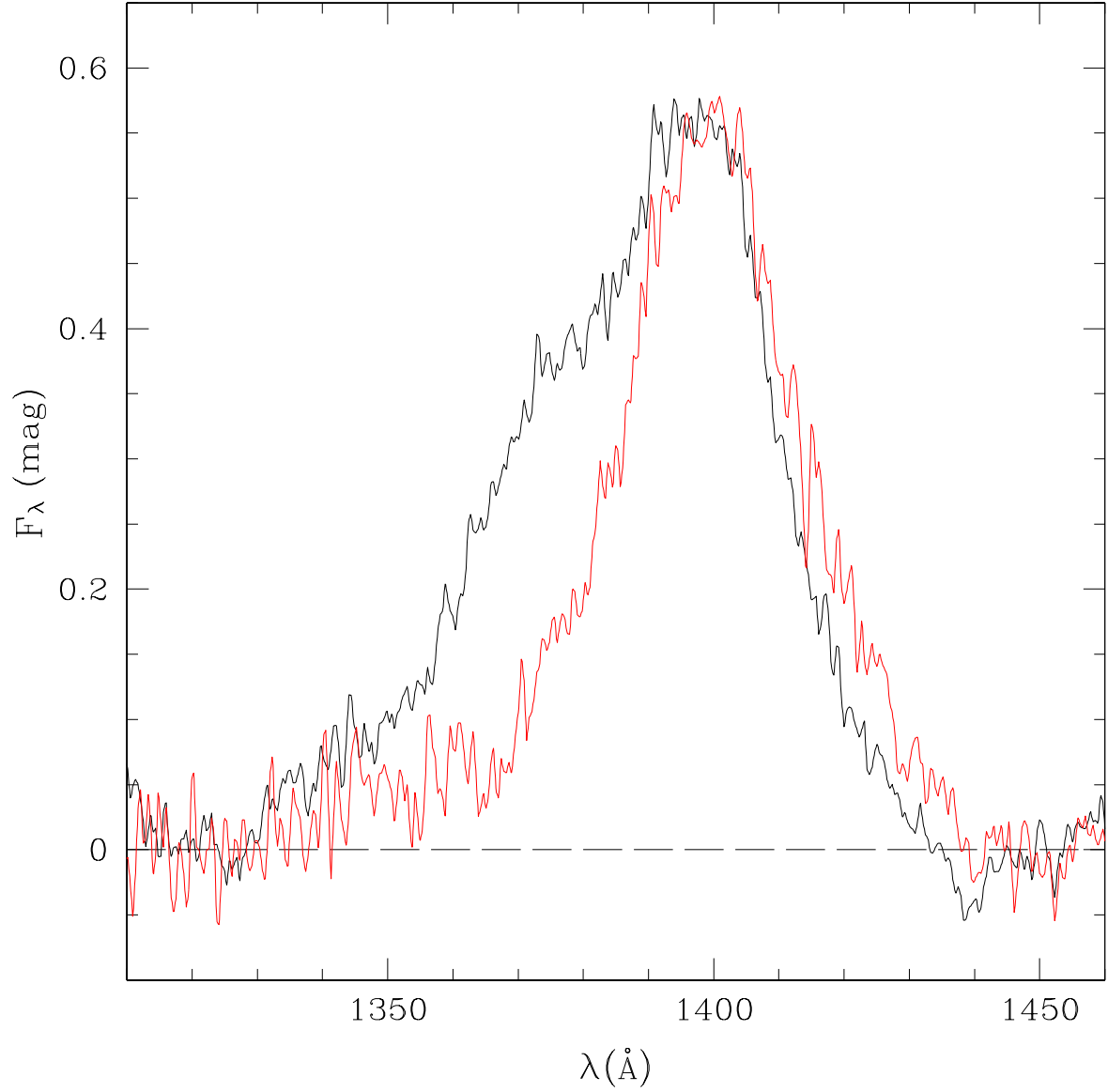}
\caption{A detailed view of the differences in the SiIV$\lambda1400$ line profiles corresponding to the A and B images of SDSS~J1004$+$4112 
  from \cite{Richards2004}.  \label{totufo}}
\end{figure}

In Mediavilla et al. (2009) we collected (from the literature) the UV, optical and
near-IR spectra shown in Figures~\ref{multiple} and \ref{multiple2} and summarized
in Tables~\ref{tbl-1} and \ref{tbl-2}.  After excluding some of the noisier spectra used in Mediavilla et al. (2009),
we are left with a sample of 18 pairs of lensed quasar images.  {We have divided the emission lines in two groups: low ionization lines\footnote{In the context of our study, we have included CIII] in the low ionization group because this emission line follows the behavior of the other low ionization lines in microlensing observations (e.g. Richards et al. 2004), reverberation mapping size estimates (e.g. Wandel et al. 1999) and line profile decompositions (Marziani et al. 2010).} (CIII]$\lambda1909$, MgII$\lambda2798$, H$\beta\lambda4861$ 
and H$\alpha\lambda6562$) and high ionization lines\footnote{We have included Ly$\alpha+$NV
in the high ionization group, since it is observed to have a similar reverberation
lag to CIV (\citealt{Clavel1991}). The Ly$\alpha$ flux could arise mainly from recombination in optically thin clouds where most of the high ionization metal lines arise (Allen et
al. 1982).}  (OVI]$\lambda1035$, Ly$\alpha$+NV$\lambda1216$, SiIV$+$OIV$\lambda1400$ and CIV$\lambda1549$). There is generally a very good match in the emission line
profiles between images.  However, there
are several cases where there are obvious differences in the line profiles (see, e.g., 
CIV in HE0435$-$1223DC, Ly$\alpha$+NV in SBS0909$+$532, and Ly$\alpha+$NV,
SiIV$+$OIV] and CIV] in SDSS J1004+4112BA). } SDSS J1004+4112 is a well-known
example (Richards
et al. 2004, G\'omez-\'Alvarez et al. 2006, Lamer et al. 2006, Abajas et al. 2007, Motta et al. 2012), 
where a blue bump appears in several high ionization emission
lines, as illustrated by the more detailed view of the SiIV$\lambda1400$ line in 
Figure~\ref{totufo}.

In order to quantify the effects of microlensing on the broad line region,
we want to isolate the effects of microlensing from those due to the large scale
macro magnification, millilensing (e.g. \citealt{Dalal2002}) and 
extinction (e.g. \citealt{Motta2002}).  We attempt this
by looking at differential flux ratios between the cores and wings of the
emission lines observed in two images
\begin{equation}
       \Delta m = (m_1-m_2)_{wings} - (m_1-m_2)_{core}.
\end{equation}
These magnitudes are constructed from the line fluxes found after
subtracting a linear model for the continuum emission underneath the line profile.
Since the line emission regions are relatively compact and 
the wavelength differences are small, this estimator certainly removes the
effects of the macro magnification, millilensing and extinction. {To see this explicitly for the macro magnification and extinction we can write the flux in magnitudes of the core (wings) of a given emission line of any of the images in a pair, $(m_{1,2})_{core,wings}$, as the intrinsic flux of the source, $(m_0)_{core,wings}$, magnified by the lens galaxy by an amount $\mu_{1,2}$, microlensed  by an amount $(\Delta \mu_{1,2})_{core,wings}$, and corrected by the extinction of this image caused by the lens galaxy, $A_{1,2}$,
\begin{equation}
       (m_{1,2})_{core,wings} = (m_0)_{core,wings} + \mu_{1,2} + A_{1,2} + (\Delta \mu_{1,2})_{core,wings}.
\end{equation}
Thus, the difference between wings and core fluxes cancels the terms corresponding to intrinsic magnification and extinction, $(\mu + A)_{1,2}$, and the difference between images cancels the intrinsic flux ratio, $(m_0)_{wings} - (m_0)_{core}$, leaving only the differential microlensing term,
\begin{equation}
       \Delta m = (m_1-m_2)_{wings} - (m_1-m_2)_{core}= (\Delta \mu_1 -\Delta \mu_2)_{wings} - (\Delta \mu_1 -\Delta \mu_2)_{core}.
\end{equation}}
{We are going to assume that the line core, {centered at the peak of the line} and defined by the velocity range 
$|\Delta v| < 850$~km/s, is little affected by microlensing compared to 
the wings, $\Delta m \sim (\Delta \mu_1 -\Delta \mu_2)_{wings}$. } Existing velocity-resolved reverberation maps (Denney et al. 2009,
2010, Bentz et al. 2010, Barth et al. 2011, Pancoast et al. 2012) all
find longer {reverberation} time delays in this velocity range, indicating that the material
in the line core generally lies at larger distances from the central 
engine.  Sluse et al. (2011) also found this in their microlensing analysis
of Q2237+0305.  Essentially, high velocity material must be close to the
central engine to have the observed Doppler shifts, while the
low velocity material is a mixture of material close to the black hole 
but moving perpendicular to the line of sight and material far from the
black hole with intrinsically low velocities.  As a result, the line core
should generically be produced  by material spread over a broader area and hence
be significantly less microlensed than the line wings.

{The microlensing effects will be little contaminated by intrinsic variability modulated by the lens time delays.  The expected continuum variability on such time scales is only of order 0.1~mag (MacLeod et al. 2010, generally, or Yonehara et al. 2008, in the
context of lenses).  The global line variability is then only 20-30\% of
the continuum variability because it is a smoothed response to the continuum, so differential (wings/core)  line variability effects should be small.
Thus, we expect these effects to represent only a modest contribution to the apparent noise.}
 
\begin{figure}[h]
\plotone{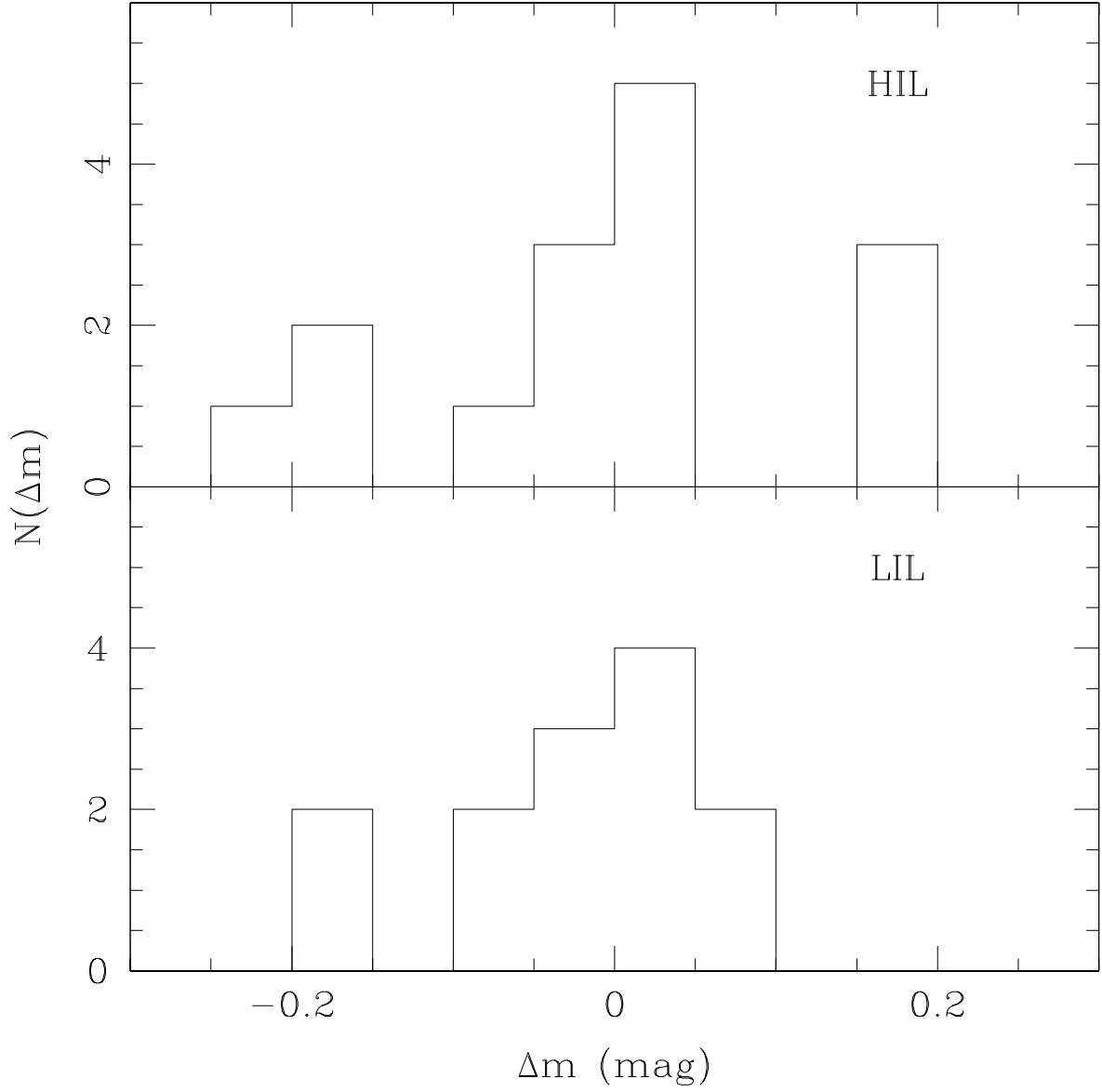}
\caption{Histograms of the microlensing magnifications, $\Delta m$, observed for the high (upper) and low (lower) ionization
  lines. \label{histo_bel}}
\end{figure}

Figure~\ref{histo_bel} shows histograms of $\Delta m$ for the low and high
ionization lines, and the values are reported in Table~\ref{tbl-2}.  The
first point to note is that even the largest microlensing effects are
relatively small, with $|\Delta m|<0.2$~mag.  The second point to note
is that more HIL (6 of 15) than LIL
(2 of 13) show significantly non-zero magnifications, $|\Delta m|\gtrsim 0.15$, given the typical 
($0.05$~mag) uncertainties (here we are counting only image pairs
showing the anomalies, not the numbers of lines showing anomalies,
so a system like SDSSJ~1004$+$4112 with multiple high ionization anomalies
is counted only once). {A binomial distribution predicts a low probability (6\%) of reproducing the HIL fraction of $|\Delta m|\gtrsim 0.15$  given the LIL
fraction.} {Qualitatively, both high
and low ionization lines are weakly microlensed but the LIL in our sample seem to be less affected by microlensing than the HIL at a $\sim 2\sigma$ level of confidence. Although the confirmation of this last result would benefit from a larger and more homogeneous sample with simultaneous observations of the HIL and LIL, there is no obvious bias in the data that would yield this result.  Moreover, 5 of the 6 image pairs that show significant microlensing of the HIL ($|\Delta m|\gtrsim 0.15$), were also observed in the LIL. The exclusion of the remaining case does not significantly affect the results of section 3.
}

\section{Constraining the Size of the Broad Line Region}


Given these estimates of the differential effects of microlensing on the
core and wings of the emission lines, we can use standard microlensing
Monte Carlo methods to estimate the size of the emission regions.  For
simplicity in a first calculation we assume that the line
core emission regions are large enough that they are effectively not
microlensed, {and simply model the luminosity profile of the region emitting the wings as a Gaussian}.
Mortonson et al. (2005) have shown that the effects of microlensing 
are largely controlled by the projected half-light area of the source,
and even with full microlensing light curves it is difficult to estimate the shape
of the emission regions (see \cite{Poindexter2010}, \cite{Blackburne2011}).
 
We use the estimates of the dimensionless surface density $\kappa$ and shear
$\gamma$ of the lens for each image from Mediavilla et al. (2009) or the
updated values for SBS~0909$+$532 from Mediavilla et al. (2011a).  We assume that
the fraction of the mass in stars is 5\% {(see, e.g., Mediavilla et al. 2009, Pooley et al. 2009, Pooley et al. 2012)}.  For a stellar mass of 
$M=1 M_\odot$, we generated square magnification patterns for each
image which were 1000~light-days across and had a 0.5~light-day pixel scale
using the Inverse Polygon Mapping algorithm (Mediavilla et al. 2006, 2011b).
The magnifications experienced by a Gaussian source of size $r_s$ ($I\propto \exp(-R^2/2 r_s^2)$)
are then found by convolving the magnification pattern with the Gaussian.  
We used a logarithmic grid of source sizes, $\ln r_s = 0.3 \times i$ for
$i=0,\cdots,17$, where $r_s$ is in units of light-days.  The source sizes can be scaled
to a different mean stellar mass, $M$, as\footnote{Lensing by stars of mass $M_\odot$ can be described in an invariant form using a characteristic length scale (Einstein radius) $\xi_0 \propto M_\odot^{1/2}$. A transformation of the mass of the stars, $M_\odot\rightarrow M$, will result in a scale change $\xi_0\rightarrow \xi_0 (M/M_\odot)^{1/2}$ that leaves invariant the dimensionless surface density, $\kappa \propto {M / \xi_0^2}$.} $r_s \propto (M/M_\odot)^{1/2}$.  {We will follow a procedure similar to that used to estimate the average size of quasar accretion disks by \cite{Jimenez2012}}.

For any pair of images, we can generate the expected magnitude differences
for a given source size by randomly drawing magnifications $m_1$ and $m_2$
from the convolved magnification pattern for the two images and taking
the difference $\Delta m = m_1 - m_2$.  The probability of observing
a magnitude difference $\Delta m_{obs,k} \pm \sigma_k $ {for image pair $k$ (averaged over the LIL or HIL, see Table \ref{tbl-2})}
given a source size $r_s$ is then
\begin{equation}
       p_k (r_s) \propto \sum_{l=1}^N \exp\left( - {1\over 2}\left({ \Delta m_l - \Delta m_{obs,k} \over \sigma_k } \right)^2\right)
\end{equation}
for $N=10^8$ random trials at each source size.  We can then estimate
an average size for either the high or low ionization lines by combining
the likelihoods
\begin{equation}
     L(r_s) = \prod p_k(r_s)
\end{equation}
for the individual {image pairs}.  Implicitly we are also drawing magnifications 
for the core but assuming they are close enough to unity to be ignored.
 
\begin{figure}[h]
\plotone{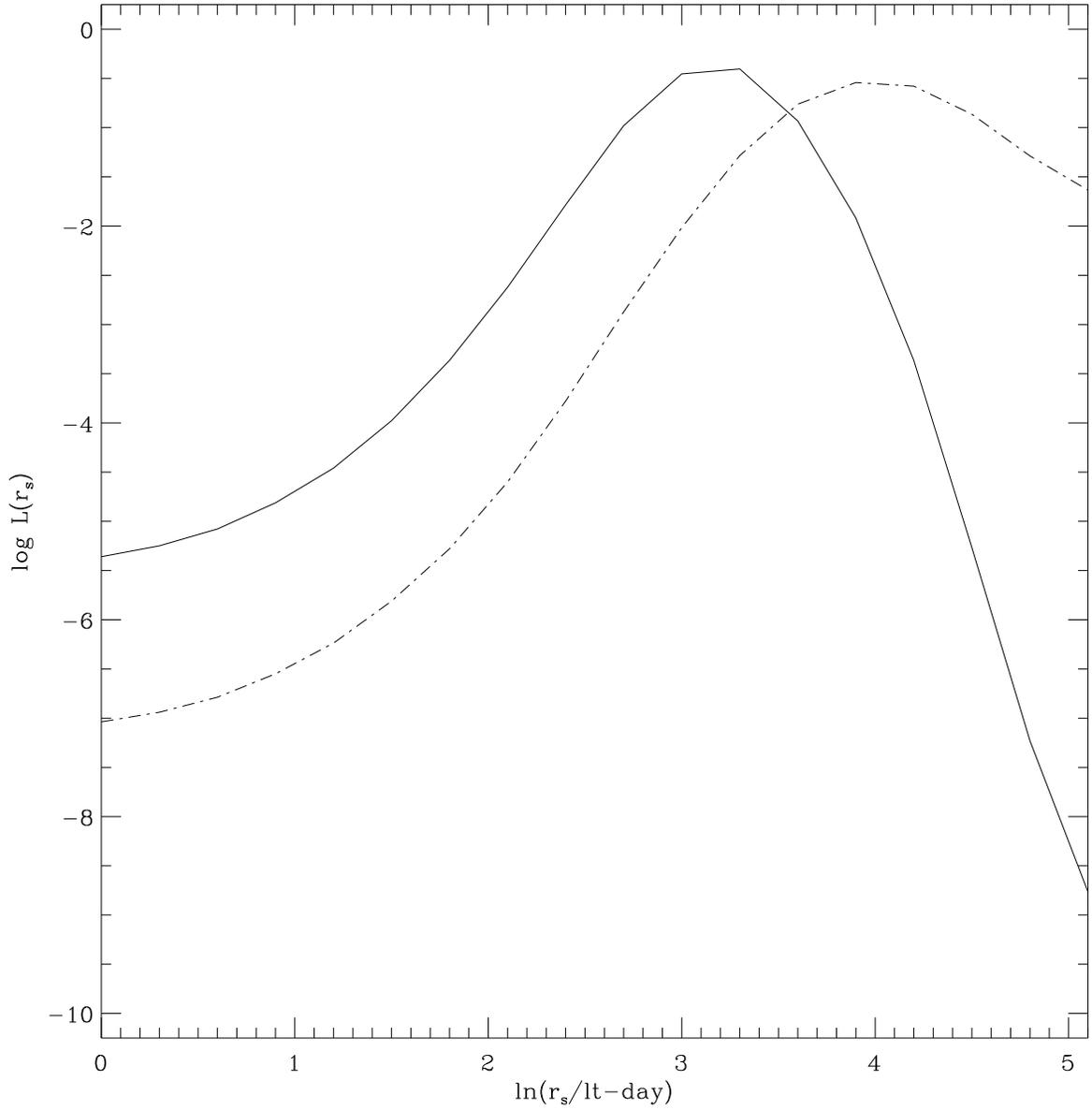}
\caption{Maximum likelihood curves for the size of the regions of high (solid) and low (dashed) ionization
  lines, respectively. \label{maxlike}}
\end{figure}

Figure \ref{maxlike} shows the resulting likelihood functions for the
high and low ionization lines.  Simply using maximum likelihood estimation, 
we find 90\% confidence estimates for the average sizes of the high and low 
ionization lines of  $r_s=24_{-15}^{+22}\sqrt{M/M_\odot}$ and $r_s=55_{-35}^{+150}\sqrt{M/M_\odot}$ light-days,
respectively {(the upper limit for the LIL was obtained using a linear extrapolation of the likelihood function)}.   {At 68\% confidence we find $r_s=24_{-8}^{+9}\sqrt{M/M_\odot}$ light-days (HIL) and $r_s=55_{-23}^{+47}\sqrt{M/M_\odot}$ light-days (LIL).} {Here we include the scaling of the inferred size with the microlenses mass, $\sqrt{M/M_\odot}$. From the likelihood functions (Fig. \ref{maxlike}), the hypothesis that the LIL and HIL have the same size is excluded at $1.8~\sigma$.}

We can make a rough estimate of the consequences of 
ignoring microlensing of the line core by raising (lowering) the magnifications to represent anti-correlated (correlated) changes
in the core relative to the line.  The effects of uncorrelated changes
will be intermediate to these limits {(more complex models explicitly including the kinematics of the emitters are explored in Appendix A)}.  {For changes of a 20\% in microlensing amplitude (from 0.8$\Delta m$ to 1.2$\Delta m$)}, the 
central sizes shift from $r_s= 20$ to $37$~light-days
for the high ionization lines and from $r_s=37$ to $120$~light-days for the
low ionization lines.

We also calculated the sizes for low ($L<2 \times 10^{44}$~ergs~s$^{-1}$) and high ($L>2 \times 10^{44}$~ergs~s$^{-1}$) luminosity sub-samples based on the
magnification-corrected luminosity estimates {at 5100\AA\ (rest frame)} from \cite{Mosquera2011}.  For the low luminosity sub-sample we find {(68\% confidence)} $r_s=16_{-8}^{+11}\sqrt{M/M_\odot}$ and  $37_{-18}^{+28}\sqrt{M/M_\odot}$~light-days for the
high and low ionization lines, while for the high luminosity sub-sample we find {(68\% confidence)} $r_s=36_{-14}^{+30}\sqrt{M/M_\odot}$ and  $r_s=299_{-103}^{+indet.}\sqrt{M/M_\odot}$~light-days. { Here we extended the grid in $r_s$ up to 400 light-days for the high luminosity, LIL case. }  While the uncertainties are
too large to accurately estimate the scaling of the size with luminosity, the changes are consistent with the $L^{1/2}$ scaling expected from simple photoionization models. 

\begin{figure}[h]
\plotone{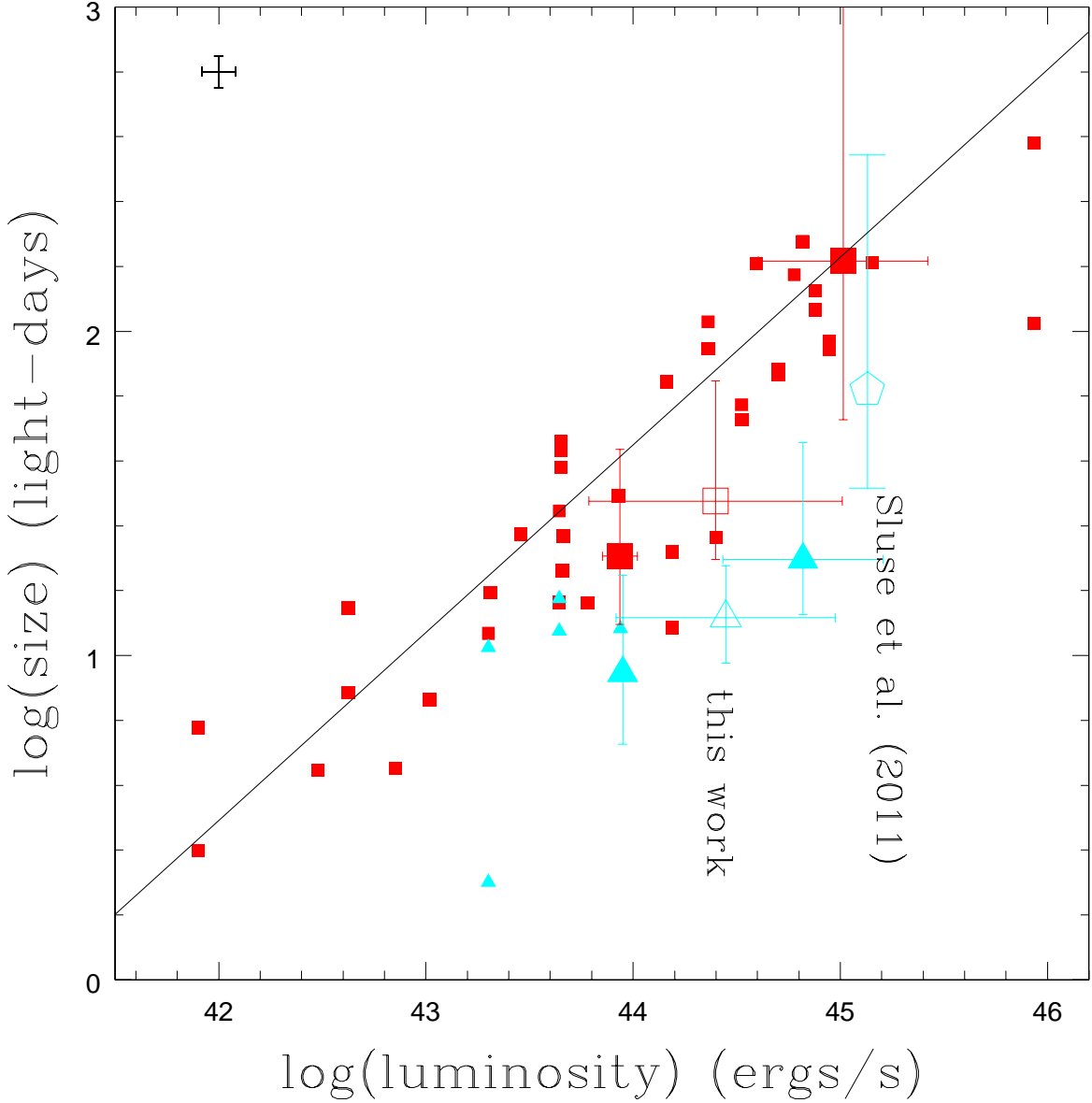}
\caption{Estimates of high (red) and low (blue) ionization broad line region sizes {(scaled to $\langle M \rangle =0.3 M_\odot$ for the mean mass and using $\pm 1\sigma$ error bars)} as a function of quasar 
  luminosity.  The present results (large triangles and squares) and the result by \cite{Sluse2011} for Q2237+0305 (large open {pentagon}) are shown using the
  magnification-corrected luminosity estimates of \cite{Mosquera2011}. The three large blue triangles (red squares) from the present work correspond to the low, total ({open symbol})
  and high luminosity subsamples defined in our data for the high (low) ionization lines (see text). The results from
  local reverberation mapping studies are shown by the small triangles (high ionization lines) and squares (low ionization lines), using the uniform estimates of the lags by
  \cite{Zu2011} and the host-corrected luminosities from \cite{Bentz2009}.  The line is the best-fit 
  correlation found by \cite{Zu2011}. The cross in the upper left corner shows the average uncertainty of the reverberation mapping lag and the variance in the source 
  luminosity during the mapping campaign. \label{fig:size}}
\end{figure}

Figure~\ref{fig:size} compares these estimates to the results from the reverberation mapping of local AGN using the uniform lag estimates by \cite{Zu2011}
and the host galaxy-corrected luminosities of \cite{Bentz2009}. In this figure 
we have scaled our estimates of $r_s$ for microlenses of $<M>=0.3 M_\odot$\footnote{This mean value is expected in typical stellar mass functions (see, e.g., Pooley et al. 2009).}. 
While the uncertainties in our
microlensing estimates are relatively large, the agreement with the reveberation  mapping results is striking.  This is clearest for the low ionization
lines which are the ones {most} easily measured in ground-based reverberation  mapping campaigns, but the offset we find between the high and low ionization lines
agrees with the offsets seen for the limited number of reverberation mapping results for high ionization lines.  We also show the estimated size of the CIV
emission region for Q2237$+$030 by \cite{Sluse2011} which shows a similar level of agreement. Because we are measuring the
size of the higher velocity line components rather than the full line, our results should be somewhat smaller than the reverberation mapping estimates for the full line.
\

\section{Conclusions}

Consistent with other recent studies (e.g. Sluse et al. 2011, 2012, Motta et al. 2012) we have found that the broad emission lines of gravitationally lensed quasars are, {in general},  weakly
microlensed.  {At a $1.8 \sigma$ level of confidence } high and low ionization lines appear to be microlensed differently, with higher magnifications observed for the higher
ionization lines.  This indicates that the emission regions associated with the high ionization lines are {probably} more compact, as would be expected from photoionization models.    If
we then make simple models of the microlensing effects, we obtain size estimates {(90\% confidence)} of $r_s=24_{-15}^{+22}\sqrt{M/M_\odot}$ and  $r_s=55_{-35}^{+150}\sqrt{M/M_\odot}$~light-days for the high and low ionization
lines.{ We have also calculated the sizes for low and high luminosity sub-samples, finding that the dependence of size with luminosity is consistent with the $L^{1/2}$ scaling expected from simple photoionization models. Our estimates also agree well with the measurements from local reveberation 
mapping studies.} These results strongly suggest that the lensed quasars can
provide an independent check of reverberation mapping results and extend them to far more distant quasars relatively economically.  Microlensing should also
be able to address the controversies about lines like CIV which have few direct reverberation mapping measurements but are crucial tools for studying the
evolution of black holes at  higher redshifts.  

With nearly 100 lensed quasars (see \citealt{Mosquera2011}) it is relatively 
easy to expand the sample and to begin making estimates of the size
as a function of luminosity or other variables.  Accurate estimates
for individual quasars will probably require spectrophotometric 
monitoring, as done by \cite{Sluse2011}.  Since the broad line
emission regions are relatively large, the time scale for the 
variability is relatively long. A significant constraint can be
gained for most of these lenses simply by obtaining one additional
spectrum to search for changes over the years that have elapsed since
many of the archival spectra we have used here were taken.  The lenses
may also be some of the better targets for reverberation studies at
higher redshifts because the time delays of the images provide
early warning of continuum flux changes and better temporal sampling
of both the line and continuum for the same investment of observing
resources.

\appendix
\section{Exploring Kinematic Models}

The problem {for analyzing more complex models including the kinematics of the emitters} is that there is no simple, generally
accepted structural model for the broad line region, and the initial results
of the velocity-resolved reverberation mapping experiments {(Denney et al. 2009,
2010, Bentz et al. 2010, Brewer et al. 2011, Doroshenko et al. 2012, Pancoast et al. 2012)} suggest that 
there may be no such common structure.  As an experiment, we constructed
a model consisting of an inner rotating disk and an outer spherical shell
which dominates the core emission.  We set the inner edge of the disk
to $r_{disk,in}=5$~light-days and left the outer edge $r_{disk,out}$
as the adjustable parameter.  For simplicity we used a constant emissivity
for the disk and a Keplerian rotation profile with an inner edge velocity
of $10^4$~km/s.  The disk has an inclination of 45 degrees. For the spherical shell we adopted fixed inner and
outer radii of $r_{sphere,in}=60$ and $r_{sphere,out}=160$~light-days
respectively.  For the shell we used a $v \propto 1/r^2$ velocity 
profile with a velocity of $5000$~km/s at the inner edge.  We normalized
the models so that the disk contributes 20\% of the flux at zero
velocity, which also results in a single peaked line profile that
resembles typical broad line profiles. We only carried out the calculations for a representative set of 
lens parameters ($\kappa_1=\gamma_1=0.45$ and $\kappa_2=\gamma_2=0.55$;
see Mediavilla et al. 2009), but we now calculate $\Delta m$
to correctly include the differential microlensing of the core and the wing.
{The final results for the outer radius of the disk component which dominates
the wings of the line profile are (68\% confidence): $r_{disk,out}=50_{-20}^{+40}$ and $r_{disk,out}= 70\pm 30$~light-days
for the high and low ionization lines, respectively. The corresponding radii enclosing half of the total disk luminosity are $r_{1/2}=r_{disk,out} /\sqrt{2}=30_{-14}^{+28}$~light-days (HIL) and $49\pm 21$~light-days (LIL). These values are in reasonable agreement with the results obtained in \S 3 without taking into account kinematics, $r_{1/2}=1.18 r_s=28_{-9}^{+11}$~light-days (HIL) and $65_{-27}^{+55}$~light-days (LIL).} {While the model is somewhat arbitrary, the similarity of the
results to the simpler analysis of \S 3 shows that it is possible to find kinematical models (probably many) that can explain the measured microlensing consistently with the hypothesis that the line core mainly arises from a region insensitive to microlensing. }

\acknowledgements

The authors thank the anonymous referee for valuable comments and suggestions. The authors are grateful to B.M. Peterson and A. Mosquera for their comments and for kindly providing data.  This research was supported by the Spanish Mi\-nis\-te\-rio de Educaci\'{o}n y Ciencia with
the grants C-CONSOLIDER AYA2007-67625-C02-02, AYA2007-67342-C03-01/03, AYA2010-21741-C03-01/02. JJV is also supported by the Junta de Andaluc\'{\i}a through the FQM-108 project.
JAM is also supported by the Generalitat Valenciana with the grant PROMETEO/2009/64. CSK is supported by NSF grant AST-1009756. VM gratefully acknowledges support from FONDECYT
through grant 1120741.

\begin{figure}[h]
\plotone{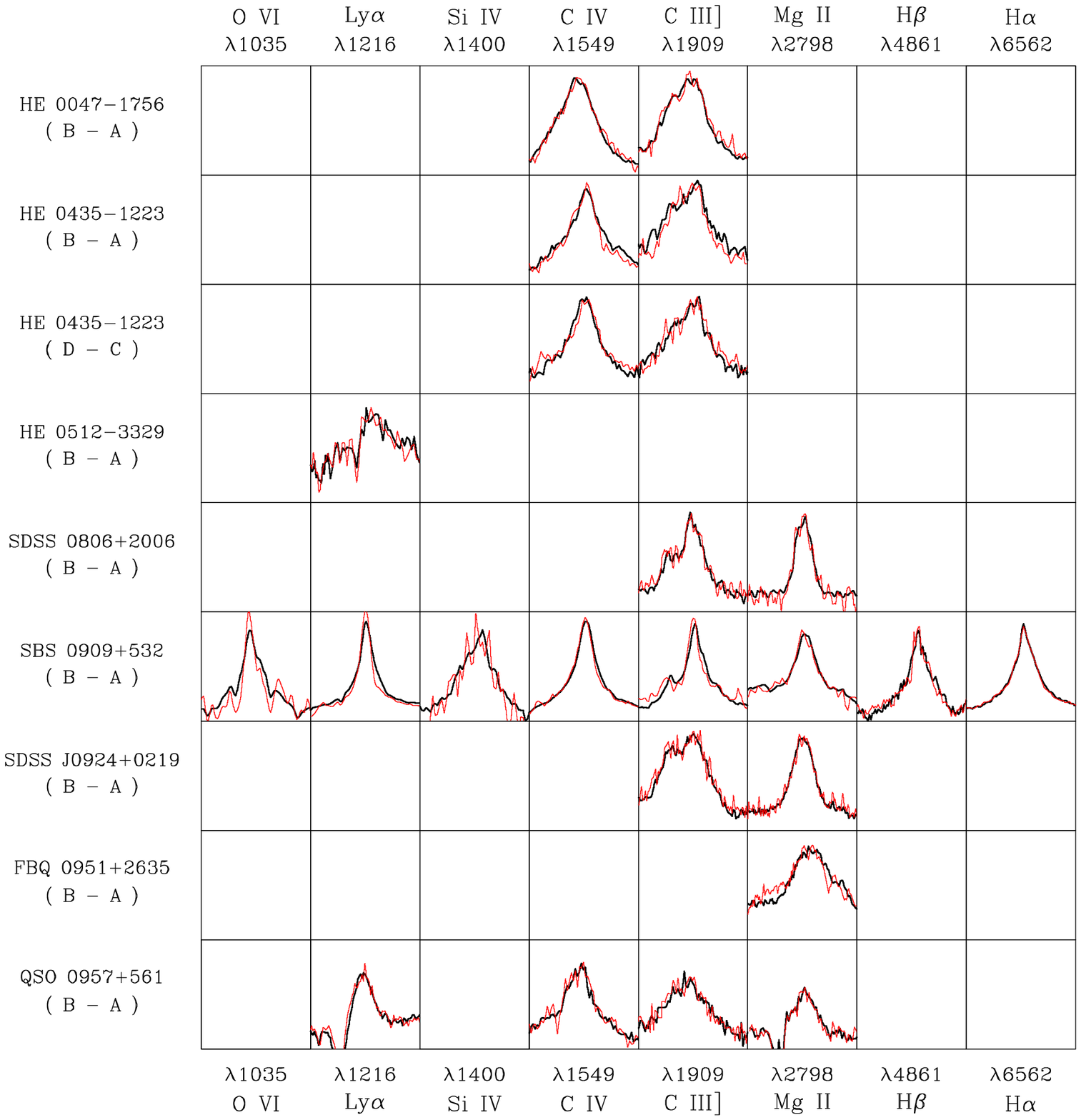}
\caption{Panels showing superpositions of emission line profiles for image pairs of several lens systems (continue in Figure \ref{multiple2}). Continuum subtracted spectra
have been scaled to match the lines. Each emission line is plotted in the ($-$6000 $\rm km\,s^{-1}$, 6000 $\rm km\,s^{-1}$) range. \label{multiple}}
\end{figure}

\begin{figure}[h]
\plotone{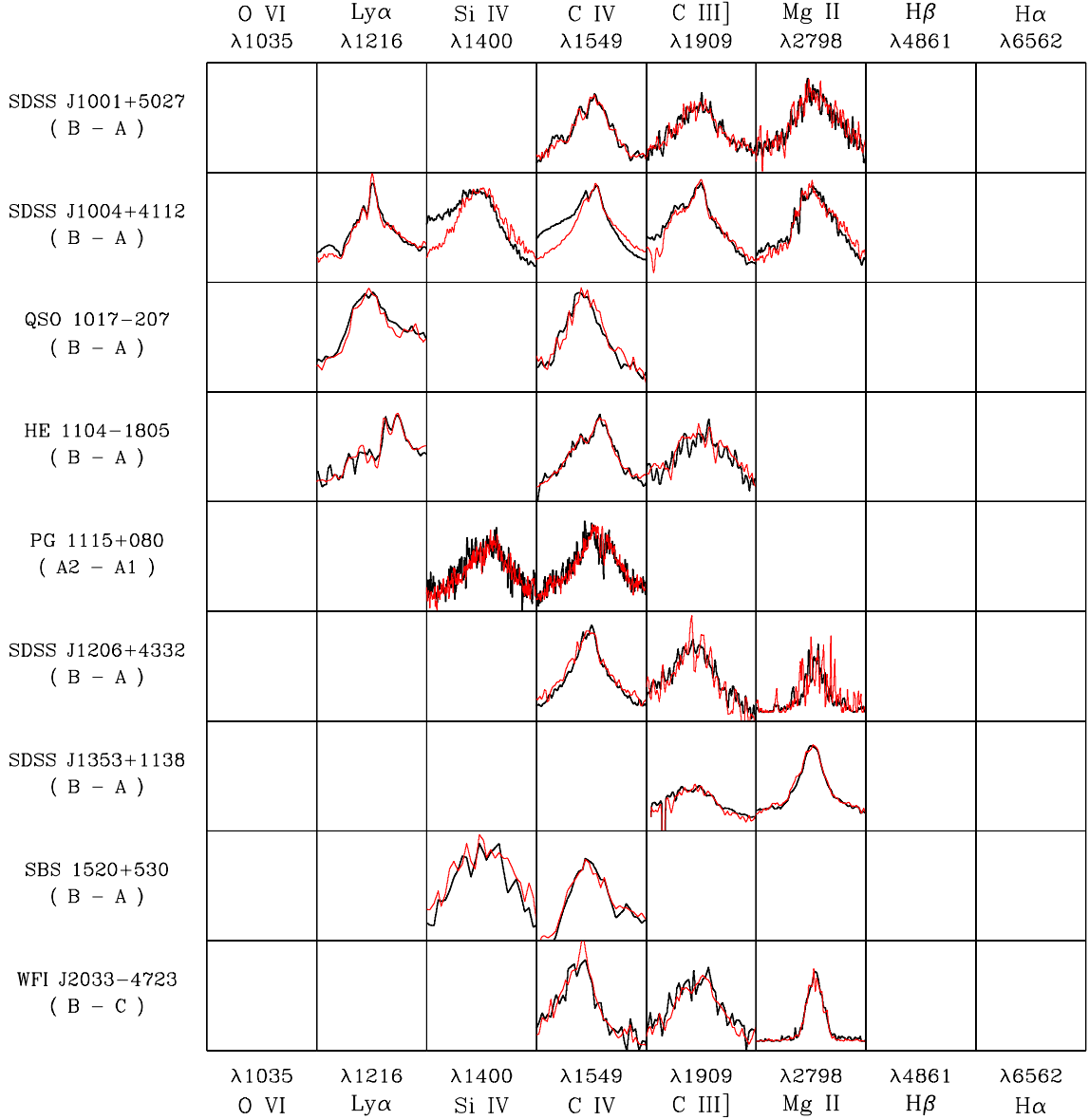}
\caption{ (continuation of Figure \ref{multiple}) Panels showing superpositions of emission line profiles for image pairs of several lens systems. Continuum subtracted spectra
have been scaled to match the lines. Each emission line is plotted in the ($-$6000 $\rm km\,s^{-1}$, 6000 $\rm km\,s^{-1}$) range.\label{multiple2}}
\end{figure}


\clearpage

\clearpage
\begin{deluxetable}{l|llccl}
\tabletypesize{\scriptsize}
\centering
\tablecaption{Lensed Quasars Sample\label{tbl-1}}
\startdata
\tableline\
Object (pair)     & $z$ & Observation Date &   Rest Wavelenght Range\tablenotemark{a}      &   Luminosity\tablenotemark{b}           & Reference \\
\tableline
HE 0047-1756  A, B  & 1.67  & 2002 Sep 04 & ( 1461  - 2547  ) & $ 2.58  \cdot 10^{  4 } $ & Wisotzki et al. 2004  \\
HE 0435-1223  A, B  & 1.689 & 2002 Sep 2-7  & ( 1413  - 2529  ) & $ 9.69  \cdot 10^{  3 } $ & Wisotzki et al. 2003  \\
HE 0435-1223  C, D  & 1.689 & 2002 Sep 2-7  & ( 1413  - 2529  ) & $ 9.69  \cdot 10^{  3 } $ & Wisotzki et al. 2003  \\
HE 0512-3329  A, B  & 1.58  & 2001 Aug 13 & ( 0775  - 2171  ) & $ 5.10  \cdot 10^{  4 } $ & Wucknitz et al. 2003  \\
SDSS 0806+2006  A, B  & 1.54  & 2005 Apr 12 & ( 1575  - 3504  ) & $ 1.09  \cdot 10^{  4 } $ & Inada et al. 2006 \\
SBS 0909+532  A, B\tablenotemark{{c}}  & 1.38  & 2005 Jan 22& ( 0957  - 2378  ) & $ 2.79  \cdot 10^{  5 } $ & Mediavilla et al. 2011  \\
& & 2004 Mar 05 & & & \\
& & 2003 Mar 07 & & & \\
& & 2001 Jan 18 & & & \\
SDSS J0924+0219   A, B\tablenotemark{{d}}  & 1.524 & 2005 Jan 14& ( 1783  - 3170  ) & $ 5.65  \cdot 10^{  3 } $ & Eigenbrod et al. 2006 \\
& &2005 Feb 01 &&&  \\
FBQ 0951+2635   A, B  & 1.24  & 1997 Feb 14 & ( 1786  - 4018  ) & $ 2.02  \cdot 10^{  5 } $ & Schechter et al. 1998  \\
QSO 0957+561  A & 1.41  & 1999 Apr 15 & ( 0913  - 4149  ) & $ 9.16  \cdot 10^{  3 } $ & Goicoechea et al. 2005  \\
QSO 0957+561  B & 1.41  & 2000 Jun 2-3  & ( 0913  - 4149  ) & $ 9.16  \cdot 10^{  3 } $ & Goicoechea et al. 2005  \\
SDSS J1001+5027   A, B  & 1.838 & 2003 Nov 20 & ( 1409  - 3136  ) & $ 4.13  \cdot 10^{  4 } $ & Oguri et al. 2005 \\
SDSS J1004+4112   A, B  & 1.732 & 2004 Jan 19 & ( 1318  - 2928  ) & $ 8.17  \cdot 10^{  3 } $ & G\'omez-\'Alvarez et al. 2006 \\
QSO 1017-207  A, B  & 2.545 & 1996 Oct 28 & ( 1016  - 1975  ) & $ 1.27  \cdot 10^{  5 } $ & Surdej et al. 1997  \\
HE 1104-1805  A, B  & 2.32  & 1993 May 11 & ( 1211  - 2846  ) & $ 1.60  \cdot 10^{  5 } $ & Wisotzki et al. 1995  \\
PG 1115+080 A1  & 1.72  & 1996 Jan 21 & ( 0846  - 1213  ) & $ 2.41  \cdot 10^{  4 } $ & Popovic, Chartas, 2005  \\
PG 1115+080 A2  & 1.72  & 1996 Jan 24 & ( 0846  - 1213  ) & $ 2.41  \cdot 10^{  4 } $ & Popovic, Chartas, 2005  \\
SDSS J1206+4332   A, B  & 1.789 & 2004 Jun 21 & ( 1362  - 3048  ) & $ 8.55  \cdot 10^{  3 } $ & Oguri et al. 2005 \\
SDSS J1353+1138   A, B  & 1.629 & 2005 Apr 12 & ( 1521  - 3385  ) & $ 1.27  \cdot 10^{  5 } $ & Inada et al. 2006 \\
SBS 1520+530  A, B  & 1.855 & 1996 Jun 12 & ( 1331  - 2452  ) & $ 4.86  \cdot 10^{  4 } $ & Chavushyan et al. 1997  \\
WFI J2033-4723  B, C  & 1.66  & 2003 Sep 15 & ( 1429  - 3008  ) & $ 8.30  \cdot 10^{  3 } $ & Morgan et al. 2004  \\
\enddata
\tablenotetext{a}{ $ $Wavelength in \AA }
\tablenotetext{b}{ $ $Luminosity corresponds to $\lambda L_\lambda$(5100\AA) in units of $ 10^{40} $ erg }
\tablenotetext{c}{ $ ${Optical, UV and near-IR spectra were obtained at different epochs}}
\tablenotetext{d}{ $ ${The spectra from the two epochs were averaged}}
\end{deluxetable}

\begin{sidewaystable}
\centering
\begin{threeparttable}


\caption{ \label{table1} Differential microlensing, $\bigtriangleup m_{core}-\bigtriangleup m_{wings}$, of the high (HIL) and low (LIL) ionization emission lines \label{tbl-2}}
\begin{tabular}{l||cccc|c|cccc|c}
\tableline\tableline
Object (pair) & $\lambda$1035 & $\lambda$1216 & $\lambda$1400 & $\lambda$1549 &$\langle$ HIL  $\rangle$& $\lambda$1909 & $\lambda$2798 &  $\lambda$4861 & $\lambda$6562 &$\langle$ LIL  $\rangle$ \\
\tableline

HE 0047--1756 (B-A)	&	-	&	-	&	-	&	+0.03	& +0.03    &	+0.03	&	-	&	-	&	-	& +0.03 \\
HE 0435--1223 (B-A) 	&	-	&	-	&	-	&	--0.21	& --0.21   &	--0.19	&	-	&	-	&	-	& --0.19 \\
HE 0435--1223 (D-C) 	&	-	&	-	&	-	&	+0.19	& +0.19    &	+0.07	&	-	&	-	&	-	& +0.07 \\
HE 0512--3329 (B-A)	&	-	&	+0.04 	&	-	&	-	& +0.04    &	-	&	-	&	-	&	-	&  -   \\
SDSS 0806+2006 (B-A)	&	-	&	-	&	-	&	-	&  -       &	+0.09	&	--0.26	&	-	&	-	& --0.09 \\
SBS 0909+532 (B-A)	&	--0.43	&	--0.23	&	-	&	--0.04	& --0.18   &	--0.01	&	--0.02	&	--0.14	&	+0.00	& --0.04 \\
SDSS J0924+0219 (B-A)	&	-	&	-	&	-	&	-	&  -	   &	+0.09	&	+0.09	&	-	&	-	& +0.09 \\
FBQ 0951+2635 (B-A)	&	-	&	-	&	-	&	-	&  -       &	-	&	+0.04	&	-	&	-	& +0.04 \\
QSO 0957+561 (B-A)	&	-	&	+0.03	&	-	&	+0.03	& +0.03    &	+0.08	&	--0.13	&	-	&	-	& --0.03 \\
SDSS J1001+5027 (B-A)	&	-	&	-	&	-	&	--0.04	& --0.04   &	+0.01	&	+0.04	&	-	&	-	& +0.02 \\
SDSS J1004+4112 (B-A)	&	-	&       --0.07  &       --0.29  &	--0.23	& --0.20   &	--0.06	&	+0.02	&	-	&	-	& --0.02 \\
QSO 1017--207 (B-A)	&	-	&	--0.08	&	-	&	+0.15	& +0.03    &	-	&	-	&	-	&	-	&  -   \\
HE 1104--1805 (B-A)	&	-	&	+0.03	&	-	&	+0.02	& +0.02    &	+0.03	&	-	&	-	&	-	& +0.03 \\
PG 1115+080 (A2-A1)	&	-	&	-	&	-0.10	&	--0.04	& --0.07   &	-	&	-	&	-	&	-	&  -   \\
SDSS J1206+4332 (A-B)	&	-	&	-	&	-	&	+0.17	& +0.17    &	--0.12	&	+0.15	&	-	&	-	& +0.01 \\
SDSS J1353+1138 (A-B)	&	-	&	-	&	-	&	-	&  -       &	--0.16	&	+0.05	&	-	&	-	& --0.06 \\
SBS 1520+530 (B-A)	&	-	&	-	&	+0.19	&	+0.16	& +0.18    &	-	&	-	&	-	&	-	&  -   \\
WFI J2033--4723 (B-C)	&	-	&	-	&	-	&	--0.05	& --0.05   &	--0.18	&	--0.14	&	-	&	-	& --0.16 \\

\tableline
\end{tabular}

\end{threeparttable}
\end{sidewaystable}

\end{document}